\documentclass[twocolumn,amsmath,amssymb]{revtex4}
\usepackage[T1]{fontenc}
\usepackage{amsmath,graphicx,amssymb,epsfig,babel,dsfont,amscd,ulem,tikz}
\usepackage{mathrsfs}
\usepackage{color}
\usepackage{placeins}
\usepackage{hyperref}
\usepackage{physics}

\renewcommand{\Re}{{\rm Re}}
\renewcommand{\Im}{{\rm Im}}
\newcommand{\rd}{{\rm d}}
\newcommand{\kb}{k_{\rm B}}

\newcommand{\ri}{{\rm i}}

\newcommand{\re}{{\rm e}}

\newcommand{\rIP}{\rm IP}
\newcommand{\rOP}{\rm OP}

\newcommand{\alphamat}{\underline{\underline{\alpha}}}
\newcommand{\chimat}{\underline{\underline{\chi}}}
\newcommand{\blockt}{\boldsymbol{T}^{-1}}

\begin{document}

\title{Surface-modes mediated long-range radiative heat transfer through a plasmonic Su-Schrieffer-Heeger chain}

\author{A. Naeimi}
\affiliation{Institut f\"{u}r Physik, Carl von Ossietzky Universit\"{a}t, 26111, Oldenburg, Germany}

\author{F. Herz}
\affiliation{Institut f\"{u}r Physik, Carl von Ossietzky Universit\"{a}t, 26111, Oldenburg, Germany}

\author{S.-A. Biehs}
\email{s.age.biehs@uol.de}
\affiliation{Institut f\"{u}r Physik, Carl von Ossietzky Universit\"{a}t, 26111, Oldenburg, Germany}

\date{\today}

\begin{abstract}
	
	We study the radiative heat transfer through a Su-Schrieffer-Heeger chain of plasmonic InSb nanoparticles in close vicinity of an InSb substrate. We show how the frequency bands of the in-plane and out-of-plane modes in the chain are deformed by the coupling to the surface waves in the InSb substrate by considering different carrier concentrations. By calculating the Zak phase we show that also in the presence of the substrate there is a topological phase transition and that topologically protected edge modes emerge for finite chains. Finally, we demonstrate the long-range heat transport along the chain due to the coupling to the surface waves of the sample {accompanied by a non-monotonic distance dependence of this effect and we show imprints of the trivial and non-trivial phase in the photonic local density of states.} We find an enhanced heat transfer in the topological non-trivial phase compared to the trivial phase due to the contribution of the edge modes.
\end{abstract}

\maketitle

%
%

\section{Introduction}\label{sec:intro}

Already in the 60's and 70's of the last century researchers have studied near-field heat radiation in pioneering experimental~\cite{Hargreaves1969,Domoto1970} and theoretical works~\cite{Polder1971} showing that the blackbody limit for radiative heat transfer can be surpassed for distances smaller than the thermal wavelength. This long-standing prediction has been verified by a series of more and more refined experiments between two planar samples separated by a vacuum gap in the last 15 years~\cite{Ottens2011,LimEtAl2015,SongEtAl2016,GhashamiEtAl2018,DeSutterEtAl2019,Fiorino2018}. There is an excellent agreement between the theory of fluctational electrodynamics and experiment down to about 25nm gap sizes~\cite{GarciaEtAl2022}. Also measurements for 7nm gap sizes with a heat flux enhancement by a factor of 18000 compared to the far-field value were reported~\cite{SalihogluEtAl2020}. Similar experiments between microscale spheres and substrates have been carried out to study the near-field regime~\cite{Narayanaswamy2008,Rousseau2009,Shen2009,Shen2010,DangEtAl2025}. In this configuration, recent experiments have verified the theoretical predictions with high precision down to only a few nanometers~\cite{GeesmannEtAl2025}. Tip-based setups claim to find a good agreement with fluctuational electrodynamics even down to angstrom distances~\cite{Kim2015,Cui2017}. In such experiments, particularly high heat fluxes can be achieved when surface waves like surface phonon polaritons (SPhPs) for dielectrics or surface plasmons (SPPs) for metals can contribute to the radiative heat transfer. This enhancement has been investigated in the above experiments as well as in experiments for thin and ultra-thin films~\cite{LimEtAl2020,MittapallyEtAl2023,SalihogluEtAl2023}, hyperbolic multilayer structures~\cite{LimEtAl2018,DuEtAl2020}, and graphene~\cite{YangEtAl, IqbalEtAl2023}, for instance. 

The surface waves cannot only increase the heat flux between two objects in the near-field regime by a direct coupling of the surface waves of the two objects but they can also serve as an efficient long-range heat flux channel. For example, when a third object like a substrate is brought in the vicinity of the two heat exchanging objects, the heat can couple into the surface waves of the substrate and enhance the heat flux between the objects~\cite{Dong2018,Messina2018,Asheichyk2017}. The length scale of this enhancement depends on the propagation length of the corresponding surface mode which is mediating the heat between the two objects similar to the Förster resonance energy transfer between two atoms coupled to a plasmonic environment~\cite{Biehs2013}. Therefore the heat transport can be long-range as studied theoretically for two nanoparticles in the vicinity of an SiC substrate~\cite{Dong2018, Messina2018}, an array of graphene strips~\cite{Zhang2019}, an SiC substrate coated with graphene~\cite{Messina2018}, a multilayer silica-graphene substrate~\cite{He2019}, an SiC sphere~\cite{Asheichyk2017}, and a planar cavity~\cite{Saaskilathi2014}. It was demonstrated that when coupling nanoparticles to a perfect metal cylinder the heat transfer can be particularly long-range~\cite{Asheichyk2022}.  Interestingly, when using a non-reciprocal substrate, the heat transfer between the two particles is not only long-range but can be tuned to prefer one direction as in a thermal diode as was shown for two InSb particles above an InSb substrate with an applied magnetic field~\cite{Ott2019, Ott2020} and for two SiC particles above a graphene sheet with an applied voltage bias~\cite{Zhang2020} as well as for two Weyl-semimetal particles above a Weyl-semimetal substrate~\cite{HuEtAl2023,Naeimi2025,Naeimi2025b}. However, the coupling to nearby structures does not necessarily enhance the heat flux, because comparing the heat flux along a free-standing particle chain with the scenario when it is brought close to another particle chain, the radiative heat transfer along the main chain is attenuated due to additional heat flux from the main chain to the other one~\cite{Luo2020}. 

In this work, we want to study theoretically the radiative heat transfer in a bipartite chain of plasmonic nanoparticles close to a plasmonic substrate supporting surface waves in the infrared. It has been shown recently that such bipartite chains of nanoparticles are an analogue of the Su-Schrieffer-Heeger (SSH) model so that they undergo a topological phase transition when the interparticle distances are altered~\cite{Ling2015, Downing2017, Downing2018, Pocock2018, Wang2020}. In particular, in the topological phase there are two isolated edge particles supporting topologically protected edge modes within the band gap of the longitudinal and transversal modes which can lead to a dominant long-range radiative heat transfer along the chain~\cite{Ott2020_2} or insulation~\cite{Nikbakht2023}. This is also true for quasi-periodic nanoparticle chains~\cite{Wang2023,Wang2023b} and 2D systems like the honeycomb lattice~\cite{Ott2021} and 2D SSH lattices~\cite{GongEtAl2024}. The coupling to edge modes can also enable long-range information exchange between distant qubits in particle chains but also in more complex networks~\cite{Lang2017}. A general review on topological heat transport can be found in Ref.~\cite{LiuEtAl2024}.  However, a combination of SSH chains with a substrate which could combine both effects -- long-range interactions due to topologically protected edge modes and due to coupling between surface and particle resonant modes -- has only been scarcely studied. Among the few examples is the possibility to use scattering type thermal microscopy to observe edge-mode signatures~\cite{Herz2022} and the possibility for a heat flux rectification along a SSH chain of SiC particles above a voltage-biased graphene coated substrate~\cite{Yang2025}.  

Our work is organized as follows: In Sec.~\ref{sec:Framework} we introduce the relations determining the band structures of a SSH chain in the presence of a substrate and the radiative heat flux through the chain. 
The emergence of topological edge modes will be elucidated by introducing the corresponding Zak phase and by studying the eigenmodes of the chain coupled to the substrate in Sec.~\ref{sec:band}. In Sec.~\ref{sec:heatflux} we will, then, investigate the radiative heat transfer along the chain and show how the edge modes and the coupling to the substrate affects the long-range heat exchange between the first and last particle. {In Sec.~\ref{sec:ldos} we discuss the local density of states in the trivial and non-trivial phase.} Finally, we summarize our main findings in Sec.~\ref{sec:conclusion}.

%
%
\section{Theoretical framework}\label{sec:Framework}

\begin{figure}
	\centering
	\includegraphics[]{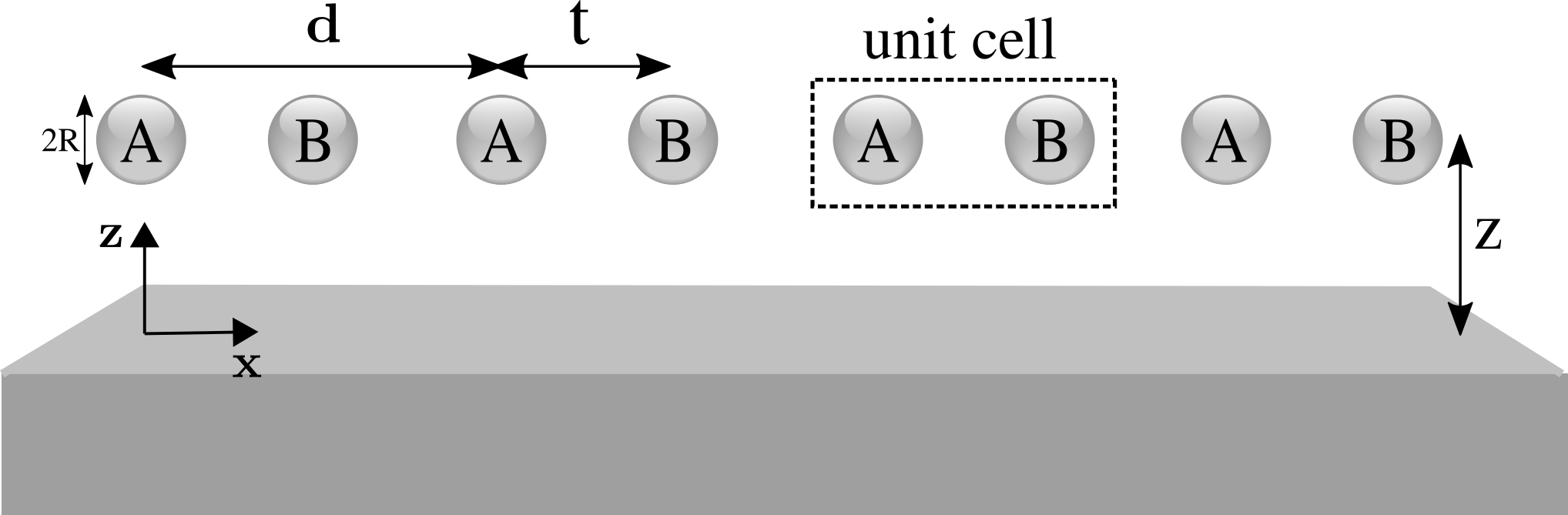}
	\caption{Sketch of the considered system. A bipartite SSH chain of nanoparticles of radius $R$ with lattice constant $d$ in each sublattice $A$ and $B$ and separation distance $t$ within each unit cell is placed at a distance $z$ in close vicinity of a semi-infinite planar substrate.}
	\label{Fig:Sketch}
\end{figure}

We consider a bipartite chain of $N$ identical spherical isotropic nanoparticles each having radius $R$ and polarizability $\alpha$ at a distance $z$ above a planar substrate as sketched in Fig.~\ref{Fig:Sketch}. The nanoparticles of the two sublattices with lattice constant $d$ are labeled by A and B. The nanoparticles A and B in a unit cell are separated by the distance $t$ and neighboring particles in adjacent unit cells by the distance $d - t$. The lattice constant $d$ and the separation distance $t$ are connected by the relation $t = \beta d / 2$ introducing the parameter $\beta$. For $\beta = 1$ all particles have the identical separation distance $t = d/2$. For $\beta < 1$ the particles in the unit cells are closer to each other than the particles of neighboring unit cells so that in consequence the dipole interaction in the unit cell is larger than the dipole interaction of particles of neighboring unit cells. For $\beta > 1$ we have the inverse situation that the dipole interaction of the nanoparticles in the unit cells is weaker than the dipole interaction of neighboring nanoparticles in different unit cells.  

\subsection{Coupled dipoles in the presence of a planar substrate}

Within the dipole approximation, the equation of dipole moment of particle $i$ interacting with the dipole moments of all other $N-1$ nanoparticles via their electromagnetic environment can be written as~\cite{ford} 
\begin{equation}
	\mathbf{p}_i = k_0^2 \sum_{j \neq i}^{N} \alpha \, \mathds{G}^{tot}(x_i,x_j) \, \mathbf{p}_j,
	\label{Eq:Eigenmodes}
\end{equation}
where $k_0 = \omega / c$ is the vacuum wave number and $c$ is the speed of light in vacuum. The total Green's function $\mathds{G}^{\rm tot}(x_i,x_j) = \mathds{G}^{0}_{ij} + \mathds{G}^{\rm sc}_{ij}$ can be decomposed into a vacuum contribution $\mathds{G}^{0}_{ij}$ and a scattering part $\mathds{G}^{sc}_{ij}$ which is due to the presence of the substrate. Note, that Eq.~\eqref{Eq:Eigenmodes} can be used to determine the eigen dipole moments and the eigenfrequencies of any configuration of nanoparticles.

In the following, we will choose to align the SSH chain along the x axis parallel to the substrate's surface as sketched in Fig.~\ref{Fig:Sketch}. The vacuum part of the Green's function can be expressed as~\cite{Messina2018} 
\begin{align}
\mathds{G}^{0}_{ij} & = G^{0}_{ij, xx} \boldsymbol{e}_x \otimes \boldsymbol{e}_x + G^{0}_{ij, \perp} (\boldsymbol{e}_y \otimes \boldsymbol{e}_y + \boldsymbol{e}_z \otimes \boldsymbol{e}_z), \\
G^{0}_{ij, xx} & = \frac{e^{{\rm i} k_0 |x_{i} - x_{j}|}}{2 \pi |x_{i} - x_{j}|} \frac{1 - {\rm i} k_0 |x_{i} - x_{j}|}{k_0^2 |x_{i} - x_{j}|^2}, \\
G^{0}_{ij, \perp} & = \frac{e^{{\rm i} k_0 |x_{i} - x_{j}|}}{4 \pi |x_{i} - x_{j}|} \frac{k_0^2 |x_{i} - x_{j}|^2 + {\rm i} k_0 |x_{i} - x_{j}| - 1}{k_0^2 |x_{i} - x_{j}|^2}.
\end{align}
The scattering part can be written as~\cite{Messina2018}
\begin{align}
	\mathds{G}^{\rm sc}_{ij} & = \sum_{\alpha = x,y,z} G^{\rm sc}_{ij, \alpha} \boldsymbol{e}_\alpha \otimes \boldsymbol{e}_\alpha \notag \\
& \quad + G^{\rm sc}_{ij, xz} \left( \boldsymbol{e}_x \otimes \boldsymbol{e}_z - \boldsymbol{e}_z \otimes \boldsymbol{e}_x \right),
\end{align}
with the components of the Green's function defined as~\cite{Messina2018}
\begin{align}
	G^{\rm sc}_{ij, xx/yy} & = {\rm i} \int_0^\infty \frac{{\rm d} k_\perp}{8 \pi} \frac{k_\perp}{k_z} e^{2 {\rm i} k_z z} \Bigl[ r_s \left( J_{0,ij} \pm J_{2,ij} \right) \notag \\
	& \quad + r_p \frac{k_z^2}{k_0^2} \left( J_{0,ij} \mp J_{2,ij} \right) \Bigr] , \label{eq:Gs1}\\
G^{\rm sc}_{ij, zz} & = {\rm i} \int_0^\infty \frac{{\rm d} k_\perp}{4 \pi} \frac{k_\perp^3}{k_z k_0^2} e^{2 {\rm i} k_z z} r_p J_{0,ij},  \label{eq:Gs2}\\
G^{\rm sc}_{ij, xz} & = \int_0^\infty \frac{{\rm d} k_\perp}{4 \pi} \frac{k_\perp^2}{k_0^2} e^{2 {\rm i} k_z z} r_p J_{1,ij}
\label{eq:Gs}
\end{align}
with the Bessel functions of the first kind $J_{n,ij} = J_n (k_\perp (x_i - x_j))$, $k_z = \sqrt{k_0^2 - k_\perp^2}$, the wavevector parallel to the surface $\mathbf{k}_\perp = (k_x, k_y)^t$, and the unit vectors $\mathbf{e}_\alpha$ in $\alpha = x,y,z$ direction. The Fresnel amplitude reflection coefficients of $s$ and $p$ polarized light are defined by
\begin{align}
r_s & = \frac{k_z - k_{z, {\rm sub}}}{k_z + k_{z, {\rm sub}}}, \\
r_p & = \frac{\varepsilon_{\rm sub} k_z - k_{z, {\rm sub}}}{\varepsilon_{\rm sub} k_z + k_{z, {\rm sub}}}
\label{eq:rp}
\end{align}
introducing $k_{z, {\rm sub}} = \sqrt{\varepsilon_{\rm sub} k_0^2 - k_\perp^2}$ inside the substrate with permittivity $\varepsilon_{\rm sub}$.

Due to the fact that the nanoparticle chain lies within the $x$-$z$ plane, the total Green's function can be decomposed into an in-plane (IP) and out-of-plane (OP) contribtion
\begin{equation}
	\mathds{G}^{\rm tot}_{ij} = \mathds{G}^{\rm tot}_{ij,\rIP} + \mathds{G}^{\rm tot}_{ij,\rOP},
	\label{Eq:Green}
\end{equation}
for which the IP part characterizes the interacting dipoles in the x-z plane and the OP one the interaction of dipoles pointing in y direction with
\begin{align}
	\mathds{G}_{ij,\rIP}^{\rm tot} &= \sum_{\alpha,\beta \in \{x,z\}} \mathds{G}_{ij,\alpha\beta} \, \mathbf{e}_\alpha \otimes \mathbf{e}_\beta, 
\label{eq:GreenIP} \\
  \mathds{G}_{ij,\rOP}^{\rm tot} &= \mathds{G}_{ij,yy} \, \mathbf{e}_y \otimes \mathbf{e}_y.
\label{eq:GreenOP}
\end{align}
Therefore, the IP Green's function in Eq.~\eqref{eq:GreenIP} describes the mixing of the longitudinal modes in direction along the SSH chain and the transversal modes perpendicular to the substrate, whereas the OP Green's function Eq.~\eqref{eq:GreenOP} describes the transversal modes in y-direction, only. Without substrate, however, only the vacuum part contributes with $\mathds{G}^{0}_{ij,xz} = \mathds{G}^{0}_{ij,zx} = 0$ and $\mathds{G}^{0}_{ij,zz}=\mathds{G}^{0}_{ij,yy}$, i.e.\ in this case the two transversal modes and the longitudinal modes are decoupled due to the symmetry in vacuum.

\subsection{Eigenvalue equation for IP and OP modes in infinitely extented chains}

To determine the band structure of the SSH chain, we consider an infinitely extended chain. Using the Bloch theorem  Eq.~\eqref{Eq:Eigenmodes} can be translated into a $2 \times 2$ block matrix equation for the $\nu= {\rm IP, OP}$ modes as
\begin{equation}
	\mathds{M}^\mathbf{\nu} \begin{pmatrix} p_{A}^{\nu} \\ p_{B}^{\nu} \end{pmatrix} = \frac{1}{\alpha}\begin{pmatrix} p_{A}^{\nu} \\ p_{B}^{\nu} \end{pmatrix}
	\label{Eq:Eigenmodes1}
\end{equation}
with block vectors $p^{\rIP}_{A/B}=(p_{x,A/B},p_{z,A/B})^{t}$, $p^{\rOP}_{A/B} = p_{y,A/B}$ and block matrix
\begin{equation}
	\mathds{M}^\mathbf{\nu} = \begin{pmatrix} M_{AA}^{\nu} & M_{AB}^{\nu} \\ M_{BA}^{\nu} & M_{BB}^{\nu} \end{pmatrix}
\end{equation}
where 
\begin{align}
	M_{AA}^{\nu} & =  M_{BB}^{\nu} = k_0^2  \sum_{j \in \mathds{Z}, j \neq 0} \mathds{G}_\nu (jd) \re^{\ri k_x j d},  
	\label{Eq:MAA} \\
	M_{AB/BA}^{\nu} &=  k_0^2 \sum_{j \in \mathds{Z}} \mathds{G}_\nu (jd \pm t) \re^{\ri k_x j d}.
	\label{Eq:MAB}
\end{align}
Note that $\mathds{G}_\nu$ replaces $\mathds{G}_{ij}^{\rm tot}$ in Eq.~\eqref{Eq:Green} with the spatial arguments $x_i-x_j = j d$ if the particles $i$ and $j$ belong to the same sublattice and $x_i - x_j = j d \pm t$ if they belong to different sublattices. The eigenvalue equation~\eqref{Eq:Eigenmodes1} now provides the eigenvectors $(p_{A}^{\nu}, p_{B}^{\nu})^t$ and eigenvalues $\alpha^{-1}$ for the coupled IP and OP modes from which the eigenfrequencies for each value of $k_x$, i.e. the frequency bands, can be derived. 

%
%
\subsection{Material properties of the nanoparticles and substrate}

In our study we choose InSb as plasmonic material for both, the SSH chain's nanoparticles and the substrate, because InSb supports surface plasmons in the infrared. The permittivity of InSb can be described by the Drude permittivity~\cite{InSb}
\begin{equation}
	\varepsilon = \varepsilon_\infty \left(1 - \frac{\omega_{\rm p}^2}{\omega(\omega+{\rm i}\Gamma)} \right).
	\label{eps1}
\end{equation}
Note that we only use the dominant electronic part of the optical response with $\epsilon_\infty = 15.68$, $\Gamma = 1\times10^{12}$ rad/s, the plasma frequency of the free carriers $\omega_{\rm p} = (\frac{ne^2}{m^*\varepsilon_0\varepsilon_\infty})^{\frac{1}{2}}$, and the effective mass $m^* = 7.29\times10^{-32}$ kg. We will choose the charge carrier concentration $n = n_{\rm p} = 1.36\times10^{19}$ cm$^{-3}$ for the nanoparticles and consider different charge carrier concentrations $n = n_{\rm sub}$ ranging from  $1.3\times10^{19}$ cm$^{-3}$ to  $1.36\times10^{19}$ cm$^{-3}$ for the substrate.

In the quasistatic limit ($R k_0 \ll 1$), the polarizability of the spherical particles in dipole approximation is given by
\begin{equation}
	\alpha = 4\pi R^3\frac{\varepsilon-1}{\varepsilon + 2}
\end{equation}
Therefore, the resonance frequency of the localized plasmonic modes in a single particle is $\omega_{\rm LP} = \omega_p \sqrt{\epsilon_\infty/(\epsilon_\infty + 2)} = 1.752\times10^{14}$ rad/s. On the other hand, the surface modes in the planar InSb substrate fulfill the dispersion relation
\begin{equation}
	k_\perp^{\rm SPP} = k_{0}\sqrt{\frac{\epsilon_{\rm sub}(\omega)}{\epsilon_{\rm sub}(\omega) + 1}}
	\label{Eq:DispSPP}
\end{equation}
where $k_\perp^{\rm SPP}$ is the wave vector of the surface wave. The surface mode resonance frequency is defined by $\Re[\epsilon_{\rm sub}(\omega_{\rm SPP})] = -1$ and $\Im[\epsilon_{\rm sub}(\omega_{\rm SPP})] \ll 1$. As we will see in the following, this resonance frequency is for all consdiered dopings higher than $\omega_{\rm LP}$. This means that the localized particle resonances can couple to surface modes in the substrate in all our considered configurations.


%
%
\subsection{Heat flux expression along the SSH chain}

Even though in some cases a Boltzmann equation approach can be used to quantify the heat flux in dipolar nanoparticle chains~\cite{PBAETAl2006,PBAETAL2008,PBAETAL2015,KathmannEtAl2018,Tervo,Tervo2} in general the framework of fluctuational electrodynamics is a better choice because it is exact within the dipole approximation and also different temperature distributions can be considered. For $N$ nanoparticles in the dipole approximation being in local equilibrium at temperatures $T_i$ in any environment at temperature $T_b$ within that framework of fluctuational electrodynamics the mean power received or emitted by particle $i$ is given by~\cite{nteilchen,Ott2020}
\begin{equation}
\begin{split}
    \mathcal{P}_{i} &= 4\Im \int_{0}^{\infty}\frac{{\rm d}\omega}{2\pi} \sum_{j=1}^{N}(\Theta_j-\Theta_b) \\
                       &\qquad \times {\rm Tr}\Big[\blockt_{ij}\chimat(\boldsymbol{DT}^{-1})_{ij}^\dagger\Big]
\end{split}
\label{pn}
\end{equation}
with the mean energy of a harmonic oscillator 
\begin{equation}
	\Theta_{i/b} = \frac{\hbar \omega}{\exp(\hbar \omega/ \kb T_{i/b}) - 1}
\end{equation}
and the generalized susceptibility
\begin{equation}
	\chimat = \Im(\alpha) \mathds{1} - k_0^2 |\alpha|^2 \frac{1}{2\ri} \bigl[ \mathds{G}^{\rm tot}_{00} -{\mathds{G}^{\rm tot}_{00}}^\dagger \bigr]
\end{equation}
which is in our configuration the same for all particles. Here the Green's function in the second term, i.e.\ the radiation correction, is the total Green's function evaluated at position $x_i = x_j = 0$. Furthermore, we have introduced the block matrices
\begin{align}
	\boldsymbol{T}_{ij} &= \delta_{ij}\mathds{1}-(1-\delta_{ij})k_0^2\alphamat\mathds{G}_{ij}^{\rm tot},  \\
	\boldsymbol{D}_{ij} &=\epsilon_0\mu_0\mathds{G}_{ij}^{\rm tot}.
\end{align}
The above expression is very general and we use it in our numerical calculations. However, for our configuration of an SSH chain we can simplify it. Neglecting the radiation correction and using the fact that IP and OP modes are decoupled, we obtain for the power received or emitted by particle $i$
\begin{equation}
	\begin{split}
		\mathcal{P}_{i}  &=  \sum_{j \neq i } \int_0^\infty \frac{\rd \omega}{2 \pi}\, {\bigl[\Theta_j - \Theta_i\bigr]} \bigl(  \mathcal{T}_{j \rightarrow i}^{\rIP} +  \mathcal{T}_{j \rightarrow i}^{\rOP}\bigr) \\
		&= \int_0^\infty \frac{\rd \omega}{2 \pi}\, ( \mathcal{P}_i^{\rIP}(\omega) + \mathcal{P}_i^{\rOP}(\omega)) .
	\end{split}
\end{equation}
with the transmission coefficient  from particle $j$ to $i$ for IP or OP direction
\begin{equation}
	\mathcal{T}_{j \rightarrow i}^{\nu} = 4 \frac{\Im(\alpha)^2}{|\alpha|^2}  \boldsymbol{T}_{\nu, ij}^{-1} \bigl(\boldsymbol{T}_{\nu}^{-1})^{\dagger}_{ij}  
	\label{Eq:TransmissionKoeff}
\end{equation}
and 
\begin{equation}
	\boldsymbol{T}_{\nu,ij} = \delta_{ij}  - (1 - \delta_{ij}) k_0^2 \, \alpha \, \mathds{G}_{\nu}^{\rm tot} (x_i - x_j).
\end{equation}
Note that here we have neglected the heat emitted to the background. This can be done, when assuming that the particle $i$ has the same temperature as the background. In the following calculations we will only set the temperature of the first particle $T_1$ to a value larger than $T_b$ and the temperatures of all other particles in the chain to $T_i = T_b$ for $i = 2, ..., N$ with the goal to evaluate the exchanged power between the first and last particle of the chain. Then Eq.~\eqref{Eq:TransmissionKoeff} quantifies the power received by each particle $i = 2,...,N$ in the chain. Keeping in mind that the expressions are based on the dipole approximation, the dipoles must be much smaller than the thermal wavelength. Furthermore, the center-to-center distance between each pair of particles and the center-to-surface distance $z$ should be larger than $3R$. In the following, we choose the sublattice constant $d = 1$ \textmu m, $0.7 \le \beta \le 1.3$, the distance to the substrate $z = 500$ nm and the particle radius $R = 100$ nm.

%
%
\section{Band structure and Edge Modes}\label{sec:band}

\begin{figure}
	\centering
	\includegraphics[]{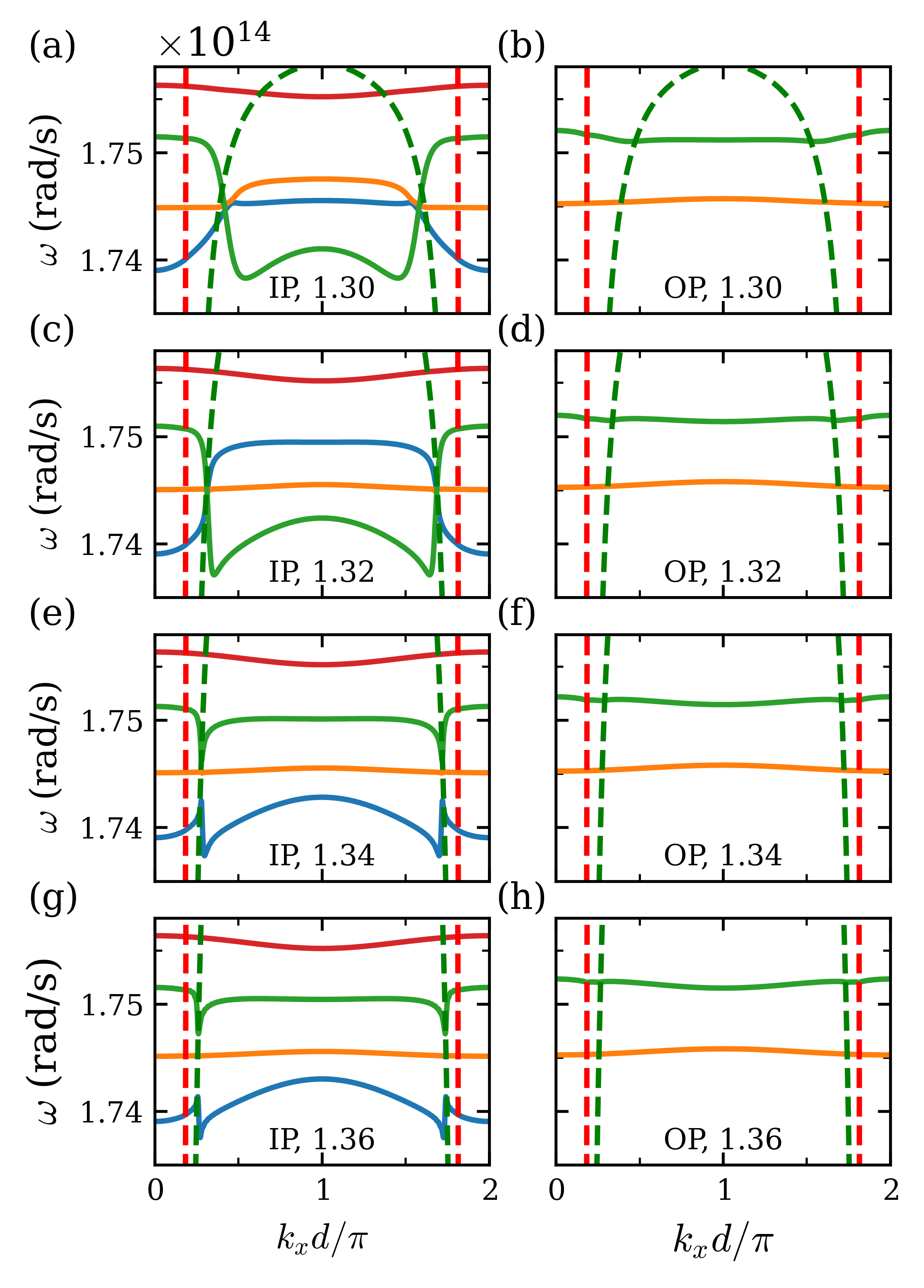}
	\caption{Real part of the eigenfrequencies of IP and OP modes form frequency bands in the first Brillouin zone. The left panels show the IP band structure and the right ones the OP band structure. The bands are evaluated for different charge carrier densities $n_{\rm sub}$ in the substrate: $n=1.30\times10^{19}$ cm$^{-3}$ for (a-b), $n=1.32\times10^{19}$ cm$^{-3}$ for (c-d), $n=1.34\times10^{19}$ cm$^{-3}$ for (e-f), and $n=1.36\times10^{19}$ cm$^{-3}$ for (g-h). The dashed lines are the light line in vacuum $k_x = \omega/c$ (red) and the surface mode dispersion relation from Eq.~(\ref{Eq:DispSPP}) (green). Here, we use $\beta = 0.7$.}
	\label{Fig:BandGap}
\end{figure}

We determine the band diagrams by solving the eigenvalue equation in Eq.~\eqref{Eq:Eigenmodes1} which is done for each $k_x$ value in the first Brillouin zone $k_x \in [0, 2 \pi/d]$. With the expresions in Eqs.~\eqref{eq:GreenIP}-\eqref{eq:GreenOP} we can obtain the bands for IP and OP separately. In Fig.~\ref{Fig:BandGap} we show the results for the real part of the eigenfrequencies for $\beta = 0.7$ of the IP modes on the left and for the OP modes on the right. Note that for an infinite chain the bands are the same for $\beta = 0.7$ and $\beta = 1.3$, because the interparticle distances are the same in both cases. As known from free standing SSH chains in vacuum~\cite{Pocock2018}, for $\beta \neq 1$ the two transversal modes and the longitudinal modes give in general two bands separated by a band gap. This bandgap closes for $\beta = 1$. With substrate we find for the OP modes still two bands separated by a band gap. However, the IP modes are mixed longitudinal and transversal modes so that here we have in general four bands as can be seen in Fig.~\ref{Fig:BandGap}.

Furthermore, it can be seen in Fig.~\ref{Fig:BandGap} that the charge carrier density $n_{\rm sub}$ of the substrate plays a crucial role for the form of the bands. Due to the coupling of the localized plasmonic resonances in the particles of the SSH chain and the surface modes in the substrate a dip in the dipolar active bands can be observed. For the OP modes this is the upper band. For the IP modes this is the upper band of the two inner bands (green and orange) which correspond to the transversal modes when the mixing of the modes can be neglected and the lower band of two outer bands (blue and red) which correspond to the longitudinal modes when the mixing of modes can be neglected. The coupling to the surface modes is relatively weak for $n_{\rm sub} = 1.36 \times 10^{19}$ cm$^{-3}$. In that case the surface modes are very close to the light line, i.e.\ the surface mode's resonance frequency $\omega_{\rm SPP}$ is relatively far away from $\omega_{\rm LP}$ and therefore also the density of surface mode states is relatively small. By decreasing  $n_{\rm sub}$ to $1.30 \times 10^{19}$ cm$^{-3}$ we increase the density of surface mode states at the particle resonance (the SPP resonance frequency $\omega_{\rm SPP} = 1.761 \times 10^{14}$ rad/s is closest to $\omega_{\rm LP}$) and the bands are stronger deformed due to a stronger coupling to the surface waves. In this case the coupling of the IP modes leads to a hybridization of the lower longitudinal and upper transversal band so that they swap parts of their bands after intersecting. To ensure in those cases that we can assign the frequencies correctly to the bands, we also considered the imaginary part of the eigenfrequencies. Real and imaginary parts for each band form continuous trajectories in the Brillouin zone which are not intersecting but surpassing each other in the additional dimension. Following these trajectories ensures a unique assignment of the real values of eigenfrequencies to their bands in the reduced representation in Fig.~\ref{Fig:BandGap}.

\begin{figure}
	\centering
	\includegraphics[width = 0.45\textwidth]{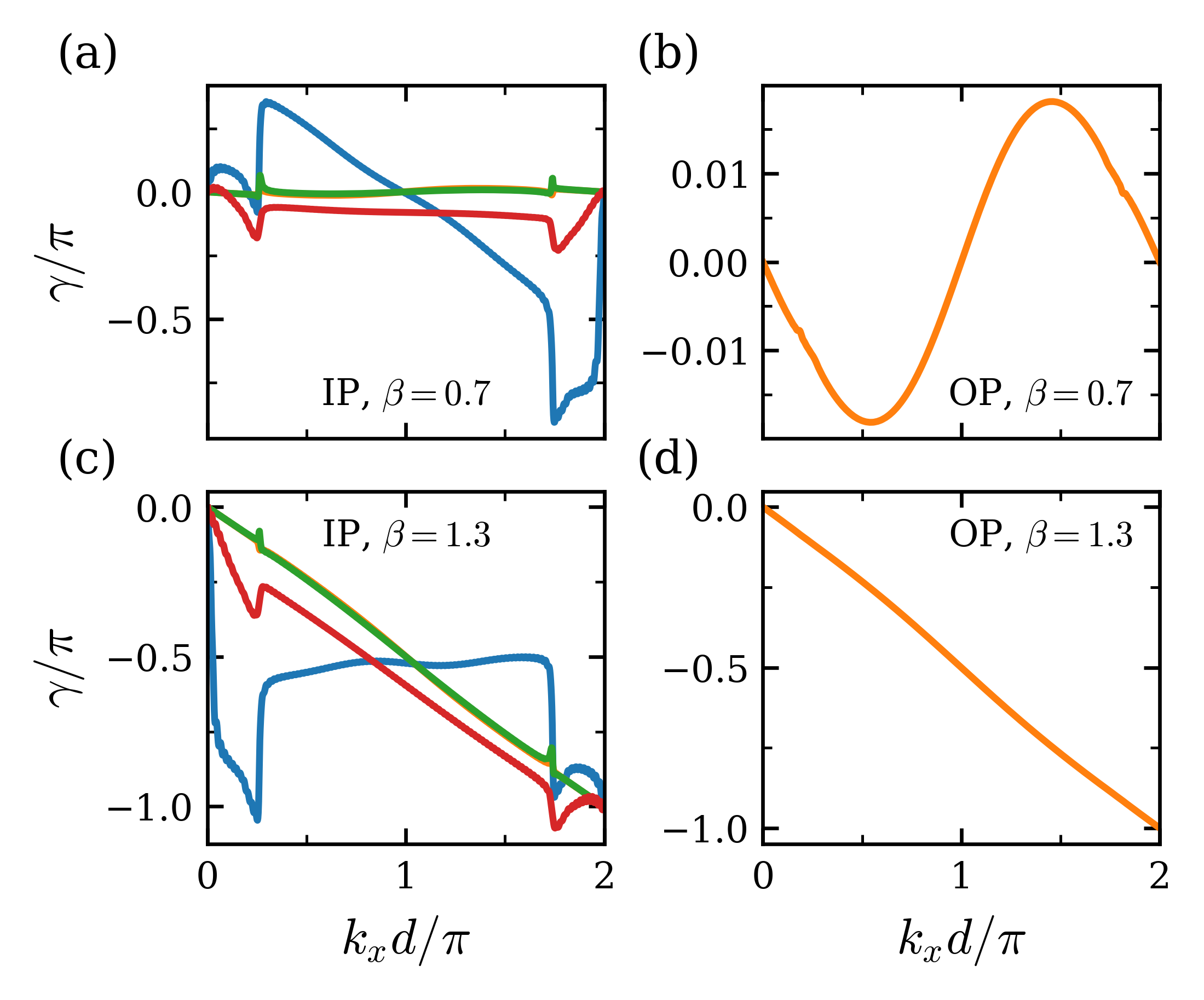}
	\caption{Accumulated Zak phase $\gamma_\nu(k_x)$ from Eq.~\eqref{Eq:Zak2} for $n=1.36 \times 10^{19}$ cm$^{-3}$. The left panels show the accumulated Zak phase for IP bands and the right ones for OP bands. The color code is identical to Fig.~\ref{Fig:BandGap}. The upper panel shows the results for the topological trivial phase with $\beta = 0.7$ and the panel at the bottom shows the result for the topological non-trivial phase $\beta = 1.3$.}
	\label{Fig:Zak}
\end{figure}

As already mentioned, for $\beta > 1.3$ we obtain the same bands as in Fig.~\ref{Fig:BandGap}. However as for the SSH chain without substrate, one can expect that $\beta < 1$ and $\beta > 1$ belong to different topological phases. To determine the topological phase, it is necessary to determine a topological invariant like the Zak phase. The Zak phase is defined by the expression~\cite{Pocock2018}
\begin{equation}
	\gamma_\nu = \ri \int_{0}^{2 \pi/d} \rd k_x\, \left( \mathbf{p}_{L}^{\nu}\right)^{\dagger} \cdot \left(\pdv{\mathbf{p}_{R}^{\nu}}{k_x}\right).
	\label{Eq:Zak}
\end{equation}
Here, $\mathbf{p}_{L/R}^{\nu}$ are the normalized biorthogonal left and right eigenvectors of Eq.~\eqref{Eq:Eigenmodes1}. The such calculated Zak phase is defined modulo $2 \pi$ as $\gamma = \gamma_0 \pm 2 n \pi$ for $n \in \mathds{Z}$ so that the value $\gamma_0 = \pi$ describes a topological non-trivial phase and $\gamma_0 = 0$ describes the trivial phase. The bulk-edge correspondence then predicts that in the topological non-trivial phase there exist edge modes for finite chains. By defining the accumulated Zak phase up to some value of $k_x$ by
\begin{equation}
	\gamma_\nu (k_x) = \ri \int_{0}^{k_x} \rd k_x'\, \left( \mathbf{p}_{L}^{\nu}\right)^{\dagger} \cdot \left(\pdv{\mathbf{p}_{R}^{\nu}}{k_x'}\right).
	\label{Eq:Zak2}
\end{equation}
we can visualize how the Zak phase is accumulated along the band in the first Brillouin zone and the value at $k_x = 2\pi/d$ then gives the full Zak phase. The accumulated Zak phase $\gamma_\nu (k_x)$ in Eq.~\eqref{Eq:Zak2} is shown in Fig.~\ref{Fig:Zak} for IP and OP modes choosing $n_{\rm sub} = 1.36 \times 10^{19}$ cm$^{-3}$. It can be seen that for the OP modes the two curves overlap clearly showing that $\gamma_0 = 0$ for both OP bands for $\beta = 0.7$ and $\gamma_0 = \pi$ for both OP bands for $\beta = 1.3$. Similarly, for the IP modes the curves of the inner bands (orange and green) overlap and $\gamma_0 = 0$ for $\beta = 0.7$ and $\gamma_0 = \pi$ for $\beta = 1.3$ for all bands. Hence, for $n_{\rm sub} = 1.36 \times 10^{19}$ cm$^{-3}$ we clearly can expect to find edge modes for $\beta = 1.3$ in the band gaps of the IP and OP modes. However for relatively low charge carrier densities like $n_{\rm sub} = 1.30 \times 10^{19}$ cm$^{-3}$ the band gaps close for the IP modes as seen in Fig.~\ref{Fig:BandGap}(a) so that the existence of edge modes is no longer guaranteed for IP modes in that case.

\begin{figure}
	\centering
	\includegraphics[width = 0.45\textwidth]{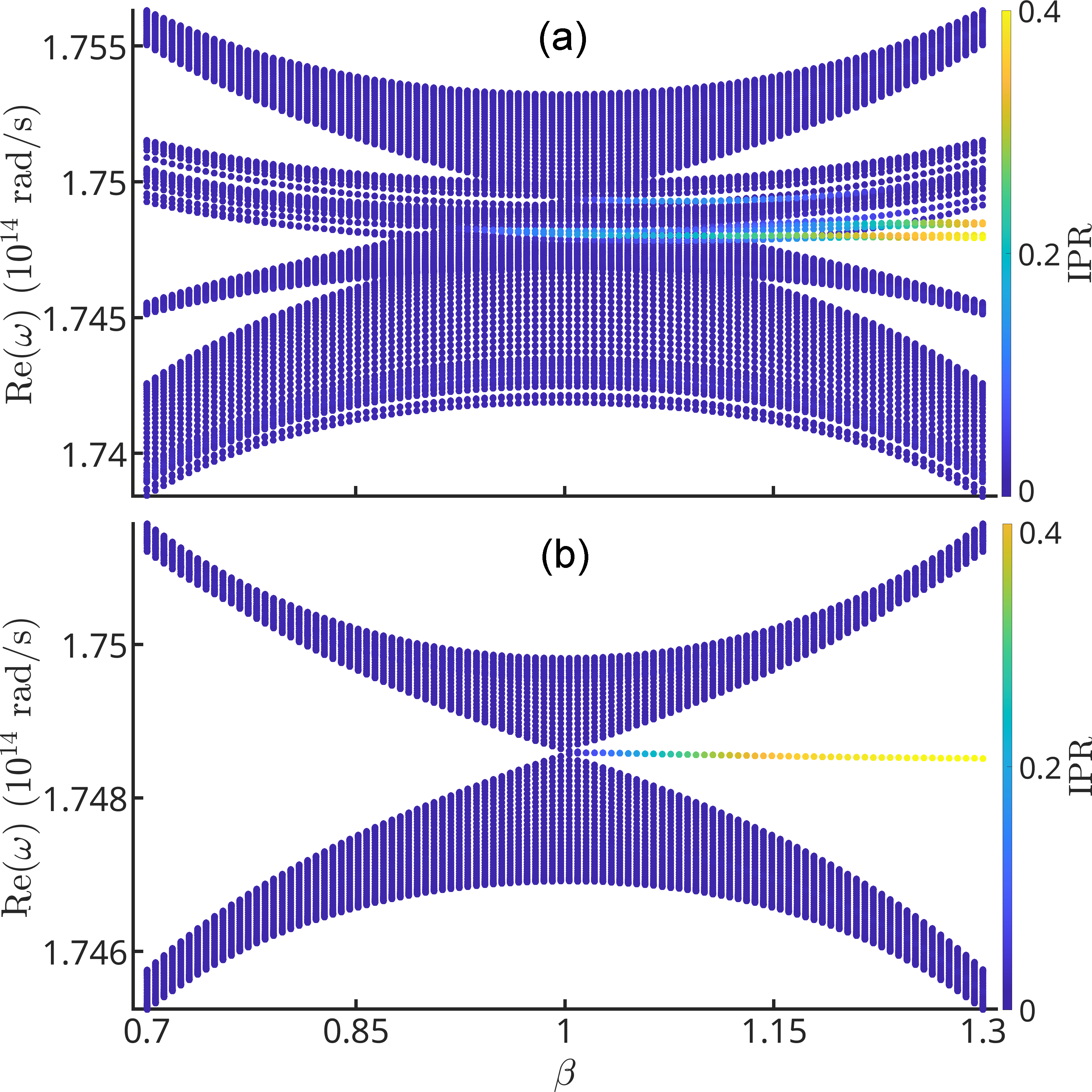}
	\caption{Real part of the eigenfrequencies for a finite chain of 60 particles and $n_{\rm sub} = 1.36 \times 10^{19}$ cm$^{-3}$ for different $\beta$ {for IP (a) and OP (b) modes. To highlight the edge modes we have added the IPR from Eq.~(\ref{Eq:IPR}) for each mode.}
	}
	\label{Fig:eigmod}
\end{figure}

To verify that there are indeed edge modes within the band gap, we show in Fig.~\ref{Fig:eigmod} the real values of the eigenfrequencies for a finite chain of 60 particles for IP and OP modes. In the case of a finite chain, the eigenfrequencies have to be calculated from Eq.~\eqref{Eq:Eigenmodes}, directly. 
{To make the difference between edge and band modes visible, we also show the inverse participation ratio (IPR) defined as~\cite{GongEtAl2024}
\begin{equation}
	\text{IPR} = \frac{\sum_{i =1}^N |\mathbf{p}_i|^4}{\bigl(\sum_{i =1}^N |\mathbf{p}_i|^2\bigr)^2}
	\label{Eq:IPR}
\end{equation}
It is derived from the eigen dipole moments $\mathbf{p}_i$ of particle $i$ and is a measure of the localisation of the modes. For band modes with $N$ evenly distributed dipole moments it would simply give $1/N$ and for a fully localized edge mode with non-zero dipole moments for the first and last particle it becomes 0.5. Therefore, small IPR on the order of $1/N$ are typical for band modes and large IPR close to 0.5 are a clear signature of highly localized edge modes.}
The resulting diagram of the eigenfrequencies as function of $\beta$ are shown in Fig.~\ref{Fig:eigmod}. For the OP modes, the two nearly degenerated topological edge modes are clearly visible in the middle of the band gap at frequency $\omega_{\rm EM,OP} \approx 1.7485\times10^{14}\,{\rm rad/s}$. For the IP modes, we can only clearly identify the four edge modes (two times two nearly degenerated edge modes) for $\beta > 1.1$. That is because the two transversal bands are too close to each other for smaller $\beta$ values so that neither a clear band gap nor edge modes are visible. For $\beta = 1.3$ the frequency of two edge modes is approximately $\omega_{\rm EM,IP} \approx 1.7479 \times 10^{14}\,{\rm rad/s}$ and the other two edge modes  $\omega_{\rm EM,IP} \approx 1.74865 \times 10^{14}\,{\rm rad/s}$.

%
%
\section{Radiative heat flux}\label{sec:heatflux}

\begin{figure}[]
	\centering
	\includegraphics[width=0.45\textwidth]{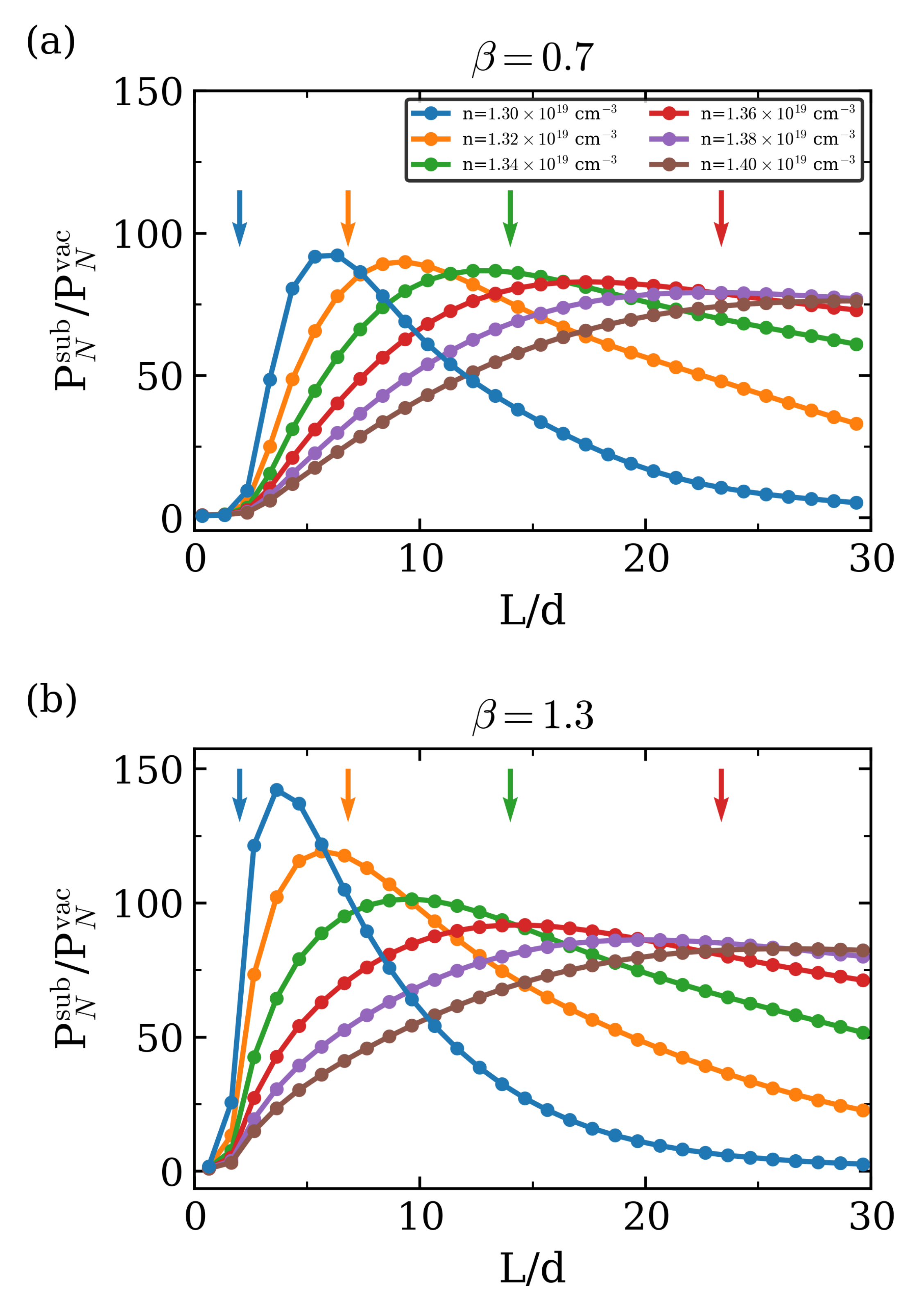}
	\caption{Total power $P^{\rm sub}_N$ received by particle $N$ in nanoparticle chains with $N = 2, \ldots, 60$ nanoparticles, i.e.\ with varying chain lengths $L$, normalized to the power $P^{\rm vac}_N$ of the same chain without substrate choosing different carrier concentrations $n_{\rm sub}$ in the substrate.}
	\label{Fig:power_L}
\end{figure}

To quantify the impact of the edge modes in the presence of a substrate on the heat flux through the chain of nanoparticles, we heat the first particle up to 310 K while the other particles and the substrate remain at 300 K. In Fig.~\ref{Fig:power_L}, we show the power $P^{\rm sub}_N$ absorbed by the last particle $N$ in the presence of the substrate normalized to the power $P^{\rm vac}_N$ in the vacuum case, i.e.\ the case without substrate, for chains with $N = 2$ to $N = 60$ nanoparticles. The power is shown as a function of the normalized chain length $L/d$ which depends on the number of particles in the chain. Furthermore, we show the results for different carrier concentrations in the substrate. In Fig.~\ref{Fig:power_L}(a) it can be clearly observed that for all cases, the presence of the substrate can increase the heat flow to the last particle up to a factor of 100 due to the contribution of the surface waves. All curves show a common feature: with increasing length of the array as the number of particles increases, the transferred power initially increases and then drops after reaching a maximum for a certain length. For the case in which the carrier concentration in the substrate is equal to the carrier concentration in the nanoparticles (red curve), this maximum occurs for a chain of 30 particles ($L \approx 15 d$) in the non-trivial case, i.e. for $\beta = 0.7$. When decreasing the carrier concentration $n_{\rm sub}$, this maximum shifts to smaller chain lengths (numbers of particles) and vice versa. The corresponding curves for the transferred power in the topological non-trivial phase for $\beta = 1.3$ are shown in Fig.~\ref{Fig:power_L}(b). The general features are the same as in the trivial phase, but the heat flux enhancement is larger (up to a factor of 120) compared to the vacuum case and the maximum is shifted to smaller chain lengths. 

\begin{table}
	\begin{tabular}{|c|c|c|c|c|c|c|}
		\hline
		$n_{\rm sub}$ $(10^{19}\,{\rm cm}^{-3})$ & 1.30  & 1.32   & 1.34      & 1.36      & 1.38      & 1.40 \\ \hline
		$\Lambda_{\rm SPP}/d $			       & 2.02 & 6.82 	& 14.02 	& 23.37 	& 34.75 	& 48.06\\ \hline 
	\end{tabular}
	\caption{ \label{table:Propagation length} Propagation length $\Lambda_{\rm SPP}(\omega_{\rm LP})$ of the surface plasmons in the substrate from Eq.~\eqref{Eq:PropLengthSPP} for different carrier concentrations $n_{\rm sub}$ in the substrate evaluated at the resonance frequency of the localized modes in the nanoparticles $\omega_{\rm LP} = 1.752\times10^{14}\,{\rm rad/s}$.}
\end{table}

The behavior observed in Fig.~\ref{Fig:power_L} can be linked to the properties of the surface waves in the substrate, and in particular to their propagation length, which can be defined as~\cite{Ott2020}
\begin{equation}
	\Lambda_{\rm SPP}(\omega) = \frac{1}{2 \Im(k_\perp^{\rm SPP}(\omega))} 
	\label{Eq:PropLengthSPP}
\end{equation}
for which the wave vector $k_\perp^{\rm SPP}(\omega)$ of the surface plasmon polariton in the substrate is defined in Eq.~\eqref{Eq:DispSPP}. A decrease in the carrier concentration $n_{\rm sub}$ red-shifts the surface plasmon resonance frequency to which the particle resonances at $\omega_{\rm LP}$ can couple and, therefore, the propagation length of the surface plasmons is reduced and vice versa. In table~\ref{table:Propagation length} we list the propagation lengths $\Lambda_{\rm SPP}$ for the different carrier concentrations in the substrate. The obtained values are depicted as arrows with the same colors like the corresponding curves for the transferred power in Fig.~\ref{Fig:power_L}. Obviously, the position of the maxima have the same trend as the propagation lenth $\Lambda_{\rm SPP}$ of the surface modes. Therefore, we can conclude that chains with a total length close to the propagation length have approximately the highest heat flux through the chain due to the contribution of the surface waves. {Consequently, for chains with lengths $L \gg \Lambda_{\rm SPP}$ the surface mode contribution declines and therefore $P_N^{\rm sub}/P^{\rm vac}_N$ declines as well.} A similar behaviour has been observed for the F\"{o}rster resonance energy transfer between two atoms mediated by the ``surface'' plasmons of a nearby sheet of graphene~\cite{Biehs2013}.

\begin{figure}[]
	\centering
	\includegraphics[width=0.45\textwidth]{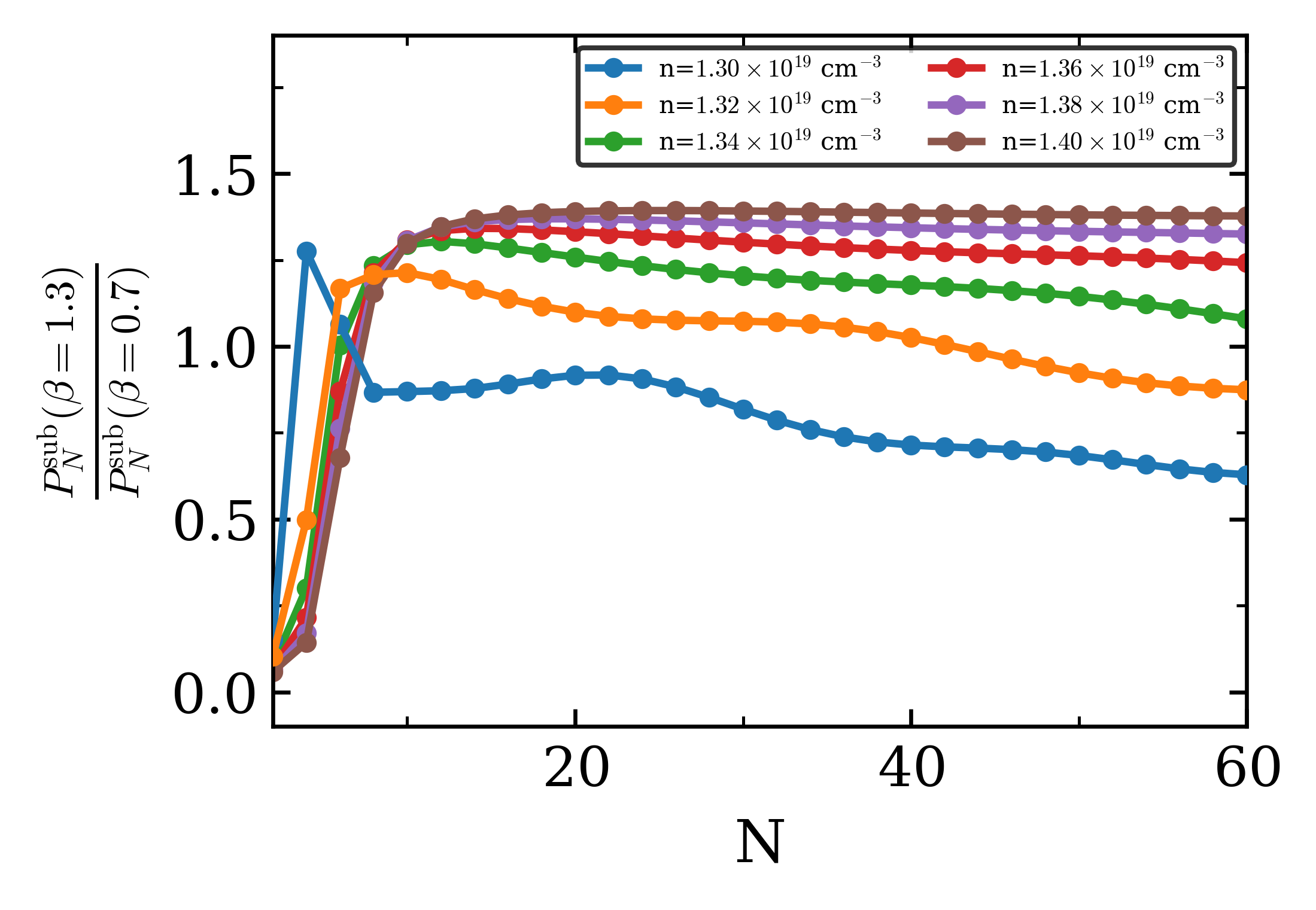}
	\caption{Power $P^{\rm sub}_N (\beta =1.3)$ received by the last particle $N$ for $\beta=1.3$ divided by the power $P^{\rm sub}_N (\beta =0.7)$ for $\beta = 0.7$ as a function of number of particles $N$ in the chain and different carrier concentrations $n_{\rm sub}$ in the substrate.}
	\label{Fig:power_N}
\end{figure}

To have a clearer picture of the impact of the edge modes in the topologically non-trivial phase, in Fig.~\ref{Fig:power_N} we show the ratio $P^{\rm sub}_N (\beta = 1.3)/P^{\rm sub}_N (\beta = 0.7)$ of the power absorbed by the last particle $N$ for $\beta = 1.3$ and for $\beta = 0.7$ as a function of the numbers of particles in the chain. It can be nicely seen that for chains with a number $N \geq 10$ and carrier concentration higher than $n_{\rm sub} = 1.34\times 10^{19}~\mathrm{cm}^{-3}$, the non-trivial phase results in larger values than the trivial phase which is a consequence of edge modes in the non-trivial phase and shows that edge modes lead to a more efficient heat flux along the chain. For the cases $n_{\rm sub} = 1.30\times  10^{19}~\mathrm{cm}^{-3}$ and $n_{\rm sub}=1.32\times 10^{19}~\mathrm{cm}^{-3}$, however, we can see that the heat flux in the non-trivial phase is dominant only for a small number of particles or short chains. In general, we find the trend that by reducing the carrier concentration $n_{\rm sub}$ of the substrate, the propagation length of the surface modes is reduced to values close to the lattice constant $d$ and, therefore, the edge mode effects can be observed for small chains only. Hence, the length scale, i.e.\ the chain length, for which an enhancement of the contribution of edge modes can be observed is also linked to the propagation length of the surface waves.

\begin{figure}[b]
	\centering
	\includegraphics[width=0.45\textwidth]{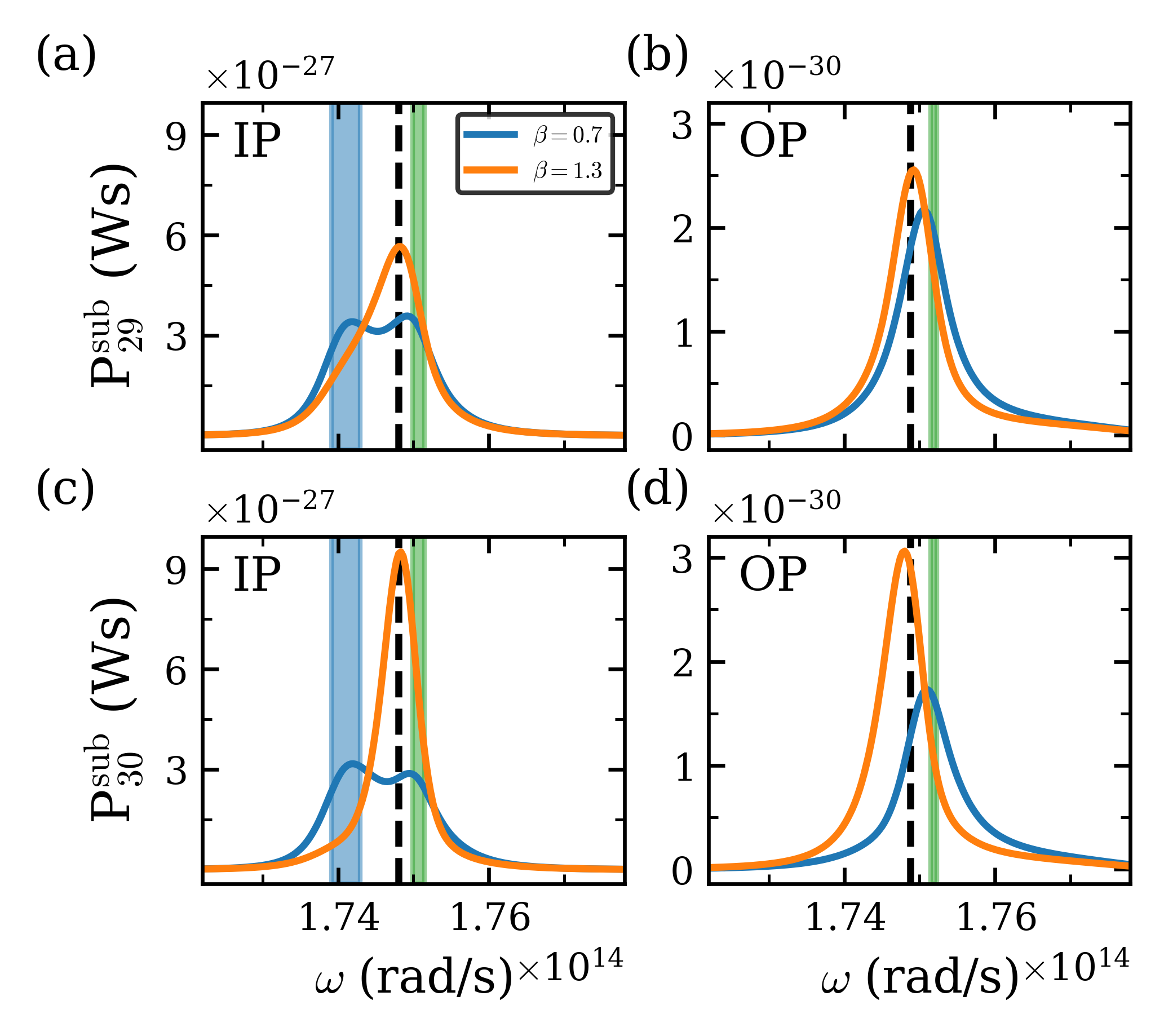}
	\caption{Spectral power $P_{29}^{\rm sub}$ and $P_{30}^{\rm sub}$ received by the 29th and 30th particle in a chain of 30 particles for $n_{\rm sub} = 1.36\times10^{19}\,{\rm cm}^{-3}$. The contribution of the IP and OP modes are shown separately. The blue and green bars in (a)/(c) are the dipole active bands determined from the real part of eigenfrequencies as in Fig.~\ref{Fig:eigmod} but for the chain of $N = 30$ nanoparticles for the IP modes and the green bar in (b)/(d) is coresponding dipole active band for the OP modes. The vertical dashed lines mark the edge mode frequencies $\omega_{\rm EM,OP} \approx 1.7491\times10^{14}\,{\rm rad/s}$ and $\omega_{\rm EM,IP} \approx 1.7479\times10^{14}\,{\rm rad/s}$, resp.}
	\label{Fig:spec29_30}
\end{figure}

In order to demonstrate the impact of the edge modes on the heat flux, we show in Fig.~\ref{Fig:spec29_30} the spectral power $P_{29}^{\rm sub}$ and $P_{30}^{\rm sub}$ absorbed by the last two nanoparticles (members of the last unit cell) in a chain of 30 particles for $\beta = 0.7$ and $\beta = 1.3$ where the substrate has the same doping level as the nanoparticles. First of all, we can see that the IP modes have a much larger impact on the heat flux than the OP modes so that the total heat flow is dominated by the IP modes. For the OP modes for $\beta = 0.7$ the power is transferred by the dipole active modes in the higher band and for $\beta = 1.3$ by the edge modes at $\omega_{\rm EM,OP} \approx 1.7491\times10^{14}\,{\rm rad/s}$ for a chain of 30 nanoparticles. For the IP modes for $\beta = 0.7$ the power is again transferred by the two dipole active bands and for $\beta = 1.3$ by the edge modes at $\omega_{\rm EM,IP} \approx 1.7479\times10^{14}\,{\rm rad/s}$ and $\omega_{\rm EM,IP} \approx 1.74865 \times 10^{14}\,{\rm rad/s}$. Comparing $P_{29}^{\rm sub}$ to $P_{30}^{\rm sub}$, we see that the edge mode contribution is higher for $P_{30}^{\rm sub}$ than for $P_{29}^{\rm sub}$. Therefore, $P_{30}^{\rm sub}$ is {approximately two times} higher than  $P_{29}^{\rm sub}$ for the IP modes demonstrating the efficient coupling of the edge modes localized at the first and last particle. Furthermore, due to the fact that the low frequency band of the IP modes dominates the heat flow for $\beta = 0.7$, the contributing modes couple to surface modes at a lower frequency than the edge modes for $\beta = 1.3$ and hence they couple to surface modes with longer propagation length. Therefore we can understand why the maximum in Fig.~\ref{Fig:power_L} is found at higher chain lengths $L/d$ in the trivial phase than in the non-trivial phase.

\begin{figure}[b]
	\centering
	\includegraphics[width=0.45\textwidth]{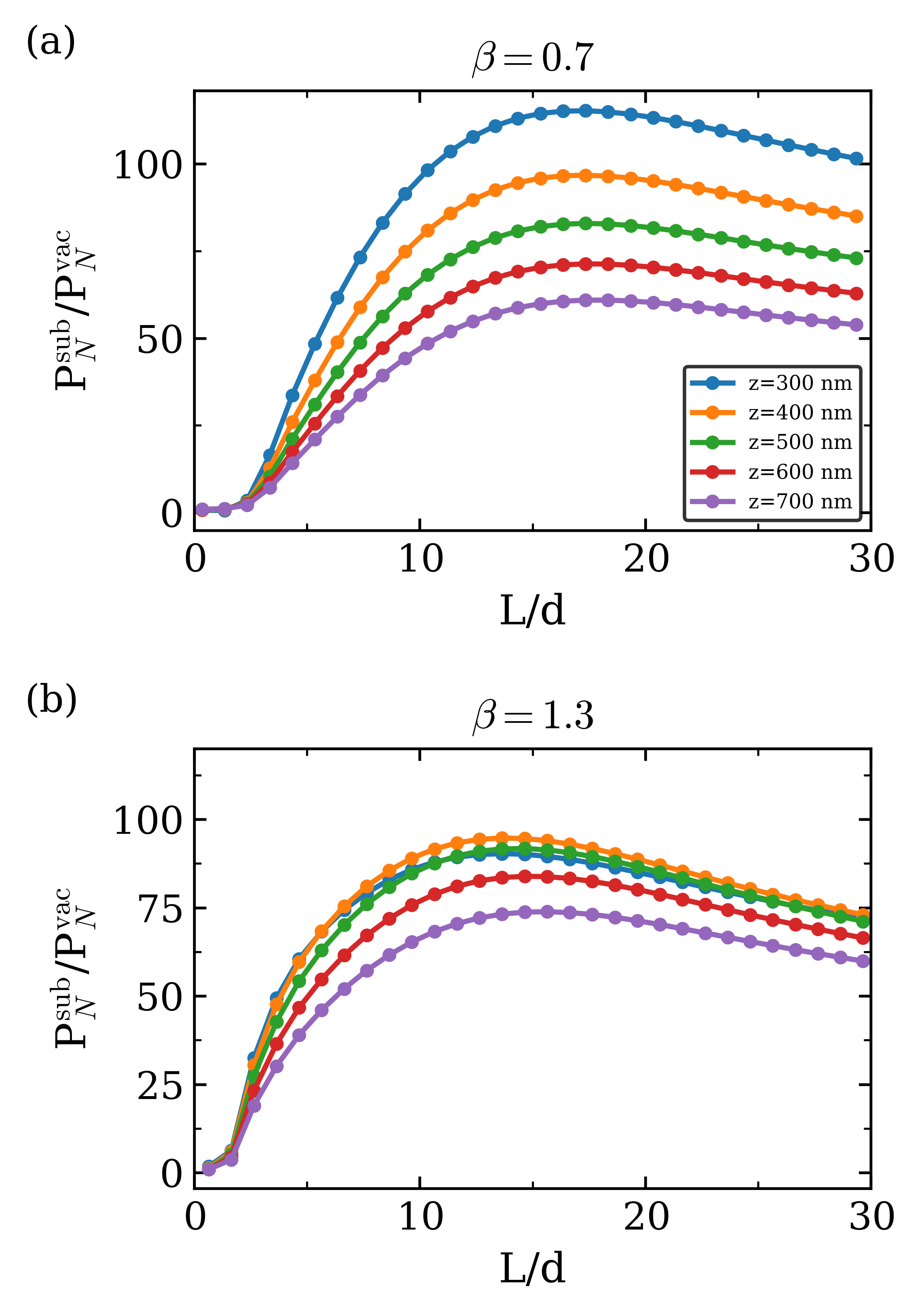}
	\caption{{Total power $P^{\rm sub}_N/P^{\rm vac}_N$ as in Fig.~\ref{Fig:power_L} for $n_{\rm sub} = 1.36\times10^{19}\,{\rm cm}^{-3}$ as a function of the chain-surface distance $z$ for $\beta = 0.7$ (a) and $\beta = 1.3$ (b).}}
	\label{Fig:PowerZdependence}
\end{figure}

\begin{figure}[b]
	\centering
	\includegraphics[width=0.45\textwidth]{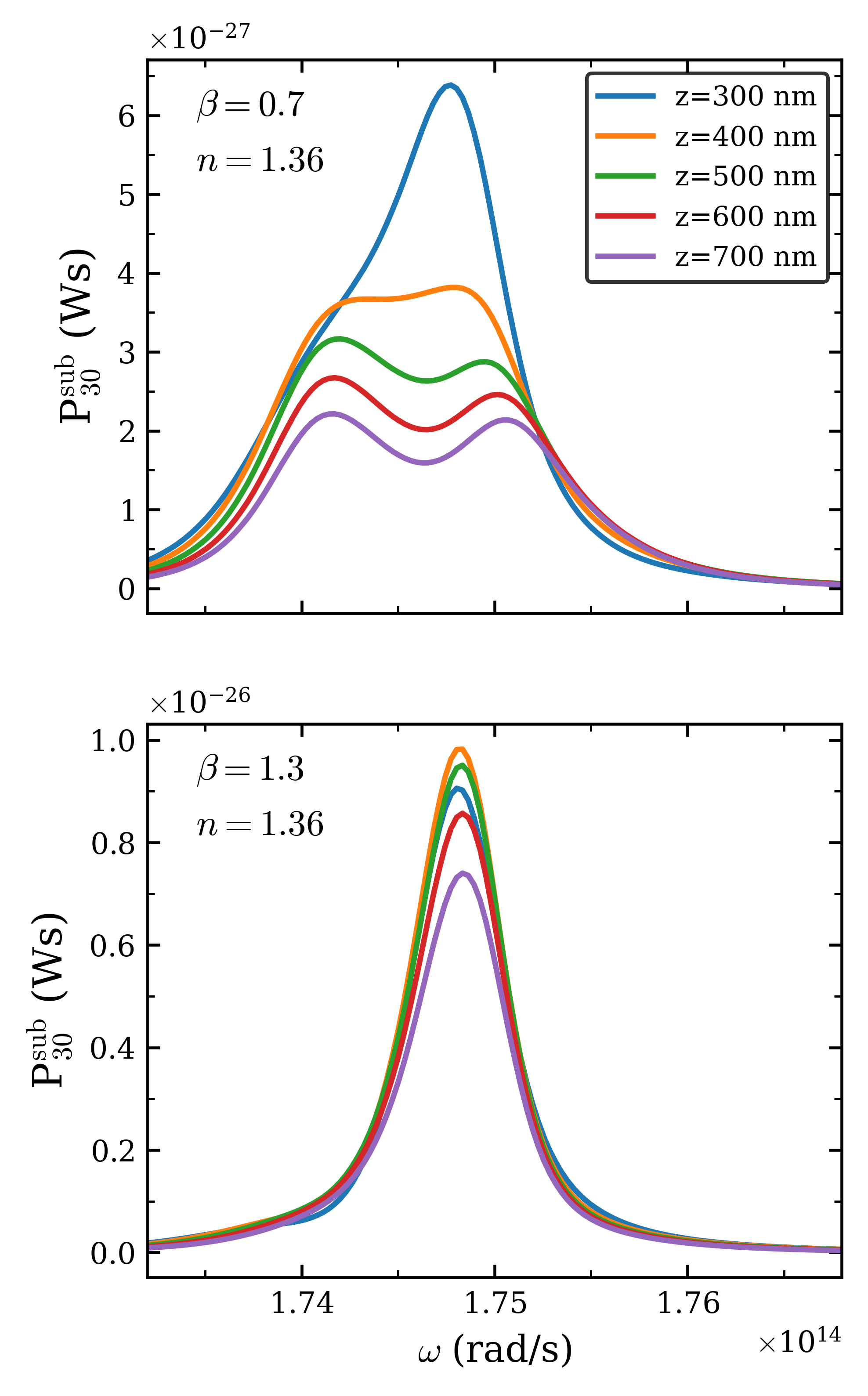}
	\caption{{Spectral power $P_{30}^{\rm sub}$ received by the last particle in a chain of 30 particles for $n_{\rm sub} = 1.36\times10^{19}\,{\rm cm}^{-3}$ for different chain-surface distances $z$. The contribution of the IP (upper panel) and OP (lower panel) modes are shown separately.}}
	\label{Fig:spec30Zdependence}
\end{figure}

\begin{figure}[b]
	\centering
	\includegraphics[width=0.45\textwidth]{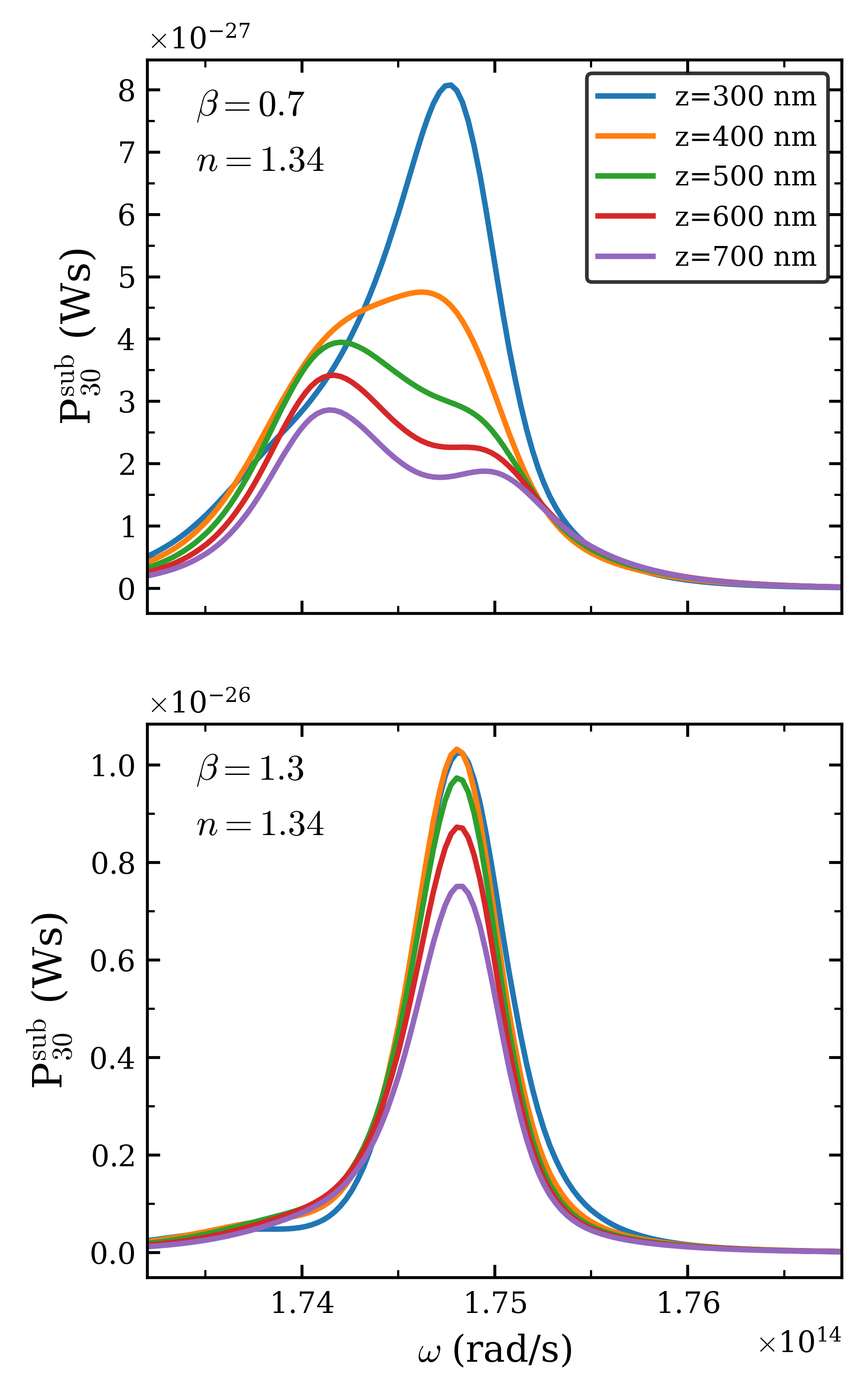}
	\caption{{As Fig.~\ref{Fig:spec30Zdependence} for $n_{\rm sub} = 1.34\times10^{19}\,{\rm cm}^{-3}$.}}
	\label{Fig:spec30ZdependenceB}
\end{figure}

\begin{figure}[tb]
	\centering
	\includegraphics[width=0.45\textwidth]{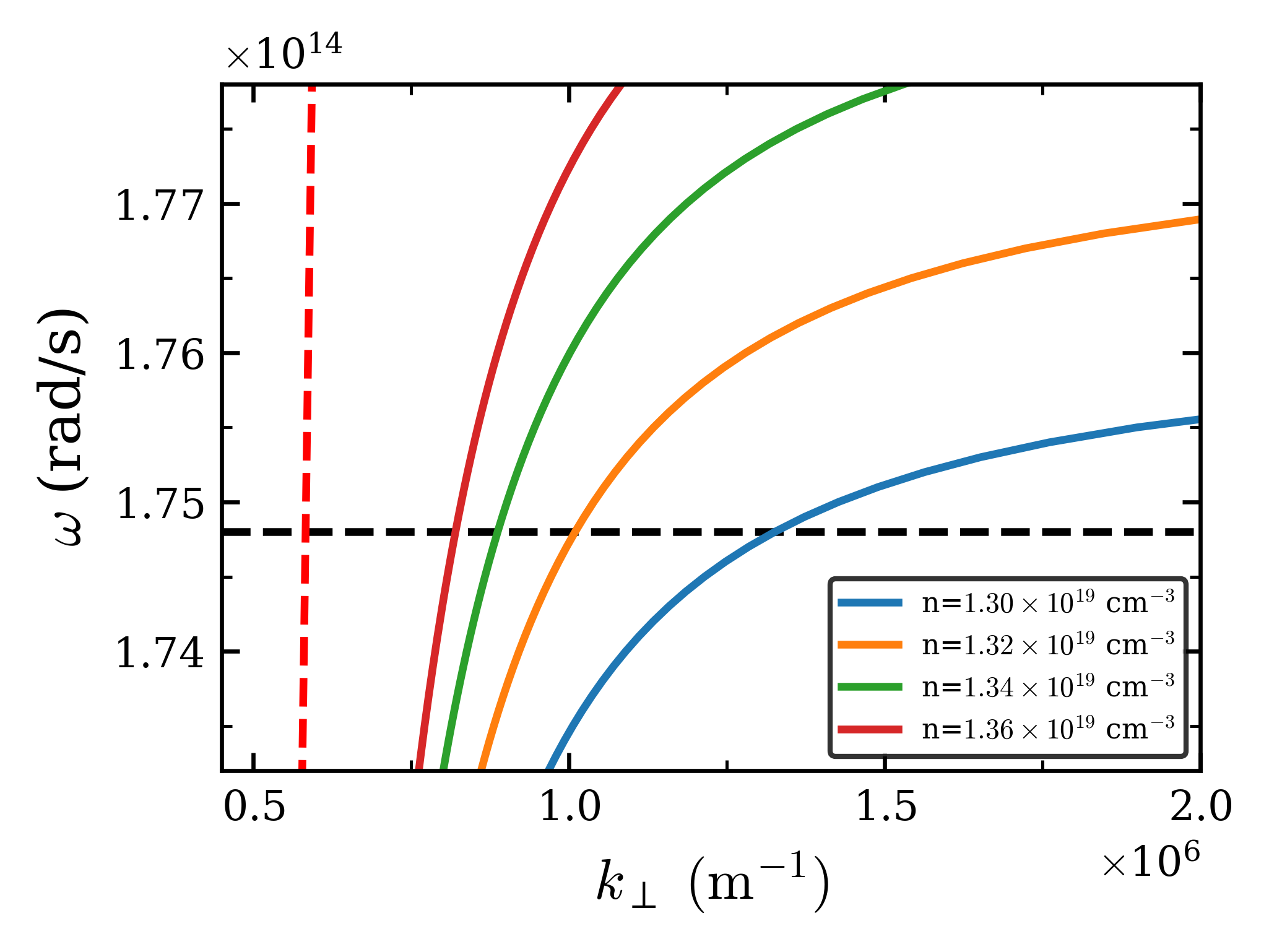}
	\caption{{Dispersion relation of the surface modes in Eq.~(\ref{Eq:DispSPP}) for different values of the electron density in the substrate. The vertical red dashed line is the light line in vacuum and the vertical black dashed line highlights a frequency close to that of the edge modes.}}
	\label{Fig:surfacemodedispersion}
\end{figure}

{Finally, we discuss the $z$-dependence of the heat transfer. In Fig.~\ref{Fig:PowerZdependence} it can be seen that for the trivial case with $\beta = 0.7$ the power $P^{\rm sub}_N$ increases when the chain is closer to the surface, indicating that the surface mode coupling increases. However, for the topological non-trival case $\beta = 1.3$ the power $P^{\rm sub}_N$ first increases when decreasing the distance to $z = 400\,{\rm nm}$ but, then, decreases for smaller distances. This indicates that the coupling strength to the surface modes in this case highly depends on the chain-surface distance $z$. In both phases, there is still the same long-range heat flux due to the coupling to the surface modes. The observed non-monotonic feature in the non-trivial phase for chain-surface distances smaller than 500 nm allows for stronger heat fluxes in the trivial phase than in non-trivial phase.}

{The same trends can be observed in the spectral power shown in Figs.~\ref{Fig:spec30Zdependence} and \ref{Fig:spec30ZdependenceB}. Furthermore, it can be seen that for $\beta = 0.7$ the relative contributions of the two bands change over distance such that for $z = 700\,{\rm nm}$ both bands contribute approximately in the same way but for distances $z < 400\,{\rm nm}$ the high frequency band dominates. Also, the location and width of the bands depend on the chain-surface distance. For $\beta = 1.3$ one can observe the non-monotonic behaviour of the maximum spectral power and a tiny red-shift of the peak frequency when $z$ is decreased being consistent with the non-monotonic behavior in Fig.~\ref{Fig:PowerZdependence}.}

{To provide an interpretation of the non-monotonic coupling behavior, we show in Fig.~\ref{Fig:surfacemodedispersion} the dispersion relation of the surface modes in the InSb substrate using Eq.~(\ref{Eq:DispSPP}). Considering energy and momentum conservation, it can be assumed getting perfect coupling between edge and surface modes if the edge mode excites a surface mode with a wavevector $k_\perp^{\rm SPP}$ whose frequency coincides with the edge mode frequency. Since the edge modes couple via the dipolar near-fields to the surface modes, the distance $z$ determines the $k_\perp$ to which the edge modes couple. As a rough estimate one can expect that $2 \Im(k_z) z \approx 1$ must be fulfilled, because the integrands in Eqs.~(\ref{eq:Gs1})-~(\ref{eq:Gs}) have a maximum for such $k_\perp$. In the quasistatic limit ($c \rightarrow \infty$) one has $\Im(k_z) = \Im(\sqrt{k_0^2 - k_\perp^2}) \approx k_\perp$ so that the rough estimate for surface mode excitation is $2 k_\perp^{\rm SPP} z \approx 1$ or $z \approx 1 /(2 k_\perp^{\rm SPP})$. From Fig.~\ref{Fig:surfacemodedispersion} one can then conclude that for $n_{sub} = 1.36\times10^{19}{\rm cm}^{-3}$ one has $k_\perp^{\rm SPP} \approx 0.825$ \textmu m$^{-1}$ and, therefore, $z \approx 1/(2 \cdot 0.825 \,) {\text{\textmu m}} = 606\,{\rm nm}$. Although this value does not exactly coincide with our numerical results, it provides a reasonably good estimate. Hence, when decreasing the distance $z$, the coupling between edge and surface modes increases until the edge modes perfectly couple to the surface modes. Further decreases in the distance lead to reduced coupling explaining the observed non-monotonuous behaviour. One can further conclude from Fig.~\ref{Fig:surfacemodedispersion} that the distance where one has the maximum coupling of the edge modes to the surface modes shifts to smaller values when the doping in the substrate decreases to $n_{sub} \leq 1.34\times10^{19}{\rm cm}^{-3}$. This trend is also found by comparing Figs.~\ref{Fig:spec30Zdependence} and \ref{Fig:spec30ZdependenceB}}

{
Interestingly, this interpretation does not work for the band modes. That is because the band modes are delocalized due to the interaction in the chain. As can be seen in the band-diagrams in Fig.~\ref{Fig:BandGap}, the dipole active bands can couple to the surface modes as long as a lattice vector coincides with $k_\perp^{\rm SPP}$. Hence, the lattice constant mainly determines whether there is a coupling or not. However, the distance of the chain to the substrate determines the local density of states, i.e.\ the coupling strength, and the value of $k_\perp^{\rm SPP}$ for best coupling. Therefore a coupling to the surface modes is guaranteed if $\pi/d > k_\perp^{\rm SPP} \approx 1/(2 z)$ in the quasistatic regime, i.e.\ for $z > d/(2 \pi)$. In our configuration with $d = 1\,\mu{m}$ this estimate results in $z > 159\,{\rm nm}$ so that in Fig.~\ref{Fig:spec30Zdependence} there is always a coupling to the surface modes via the lattice for all shown distances for which the dipole model is valid. The monotonic increase in coupling of the band modes for small distances $z$ can, then, be attributed to the increased local density of states. That means one can have a simple understanding of the distance dependent coupling strength for both the band and edge modes. }

%
%
{
	\section{Local Density of States}\label{sec:ldos}}

{
The photonic local density of states (LDOS) is another quantity which allows us to study the difference between edge and band modes more detailed and is related to other measurable quantities like the emission rates of atoms and molecules or the signal of scattering type microscopes, for instance.
The full LDOS for coupled nanoparticle-surface systems was derived in Ref.~\cite{Ott2021}. In our setup the electrical fields dominate due to the interaction of the surface modes with the resonances of the nanoparticles. This allows us to focus on the electric LDOS given by~\cite{Ott2021}
\begin{equation}
\begin{split}
	D^{\rm E}(\omega) &= \frac{k_0}{\pi c} {\rm ImTr} \bigl\{\mathds{G}_{00}^{E}\mathds{G}_{00}^{E}\big\} \\ 
	                  &\quad+ \frac{k_0^3}{\pi c}\sum_{i,j = 1}^N{\rm ImTr}\Big\{\alpha\mathds{G}_{0i}^{E}\mathds{G}_{0j}^{E}\Big\} \\
                          &\quad + \frac{k_0^3}{\pi c}\sum_{i,j = 1}^N{\rm ImTr}\Big\{\alpha\mathds{G}_{0i}^{E}\bigl(\boldsymbol{T}_{ij}^{-1} - \mathds{1}\bigr)\mathds{G}_{0j}^{E}\Big\}.
\end{split}
	\label{Eq:eLDOS}
\end{equation}
Here, the first term is the contribution of the surface, the second term describes the contribution of the nanoparticles, and the third term contains the interaction of the nanoparticles with the surface. This interaction term
\begin{equation}
	D^{\rm E}_{\rm I}(\omega) = \frac{k_0^3}{\pi c}\sum_{i,j = 1}^N{\rm ImTr}\Big\{\alpha\mathds{G}_{0i}^{E}\bigl(\boldsymbol{T}_{ij}^{-1} - \mathds{1}\bigr)\mathds{G}_{0j}^{E}\Big\}
\label{Eq:eLDOSI}
\end{equation}
can have either positive or negative values whereas the other two terms are purely positive describing the LDOS of the decoupled system.
}

\begin{figure}[tb]
	\centering
	\includegraphics[width=0.45\textwidth]{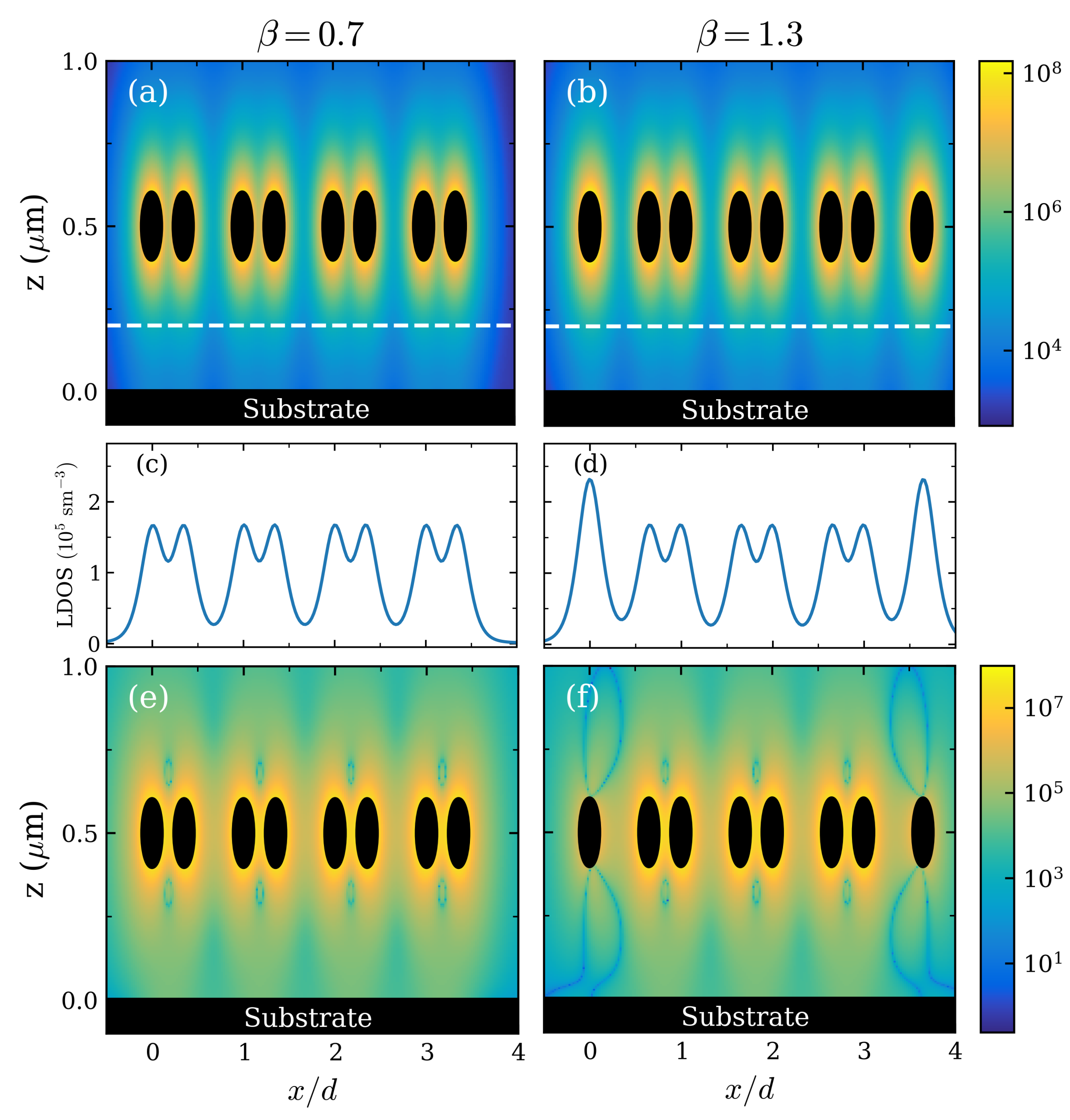}
	\caption{{Electric LDOS for $\beta = 0.7$ (left panels) and $\beta = 1.3$ (right panels). (a), (b) full LDOS $D^{\rm E}(\omega)$ in (s/m$^{3}$) in the two-dimensional plane of the nanoparticle chain. (c), (d) line scans of the full LDOS in (s/m$^{3}$) along the dashed white line in (a) and (b) i.e.\ for $z=200$ nm. (e), (f) modulus of the interaction LDOS $|D^{\rm E}_{\rm I}(\omega)|$ in (s/m$^{3}$) in the two-dimensional plane of the nanoparticle chain.}}
	\label{Fig:LDOS}
\end{figure}

{
	In Fig.~\ref{Fig:LDOS}(a), (b) we show the full LDOS $D^{\rm E}(\omega)$ for $N = 8$ nanoparticles from Eq.~(\ref{Eq:eLDOS}) evaluated at the corresponding edge mode resonance frequency $\omega_{\rm em} \approx 1.7479\times10^{14}\,{\rm rad/s}$ of the IP modes for $\beta = 0.7$ and $\beta = 1.3$. The difference between the trivial and non-trivial phase is barely visible in these 2D plots but can be nicely seen in the line scans of the LDOS at $z = 200\,{\rm nm}$ in Fig.~\ref{Fig:LDOS}(c),(d). At the position of the isolated particles there is an increased LDOS in the non-trivial phase compared to the trivial phase demonstrating that the edge modes increase the LDOS in the non-trivial phase similar to what has been observed for SSH chains without substrate in Ref.~\cite{Ott2021}. In Fig.~\ref{Fig:LDOS}(e),(f) we also show the modulus of the interaction term $D^{\rm E}_{\rm I}(\omega)$ from Eq.~(\ref{Eq:eLDOSI}) for the same configuration. Actually, this interaction term is mainly negative, i.e. due to the interaction the LDOS around the nanoparticles is reduced. For the edge particles and $\beta = 1.3$, this reduction is less strong than for $\beta = 0.7$ so that, in total, the full LDOS is larger in the non-trivial phase. Furthermore, for $\beta = 1.3$ lobes can be seen above and below the edge particles. There, the interaction LDOS $D^{\rm E}_{\rm I}(\omega)$ is positive, i.e.\ the LDOS is enhanced. The lobes below the particles are deformed due to the interaction with the surface modes providing a visible signature of the enhanced coupling in the non-trivial phase compared to the trivial phase. Hence, the full LDOS indicates the enhancement close to the edge modes and the interaction LDOS shows the interaction with the surface.}

%
%
\section{Conclusion}\label{sec:conclusion}

To summarize, we have investigated the radiative heat transfer through a plasmonic SSH chain of InSb nanoparticles
coupled to a planar InSb substrate within the framework of fluctuational electrodynamics using the dipole approximation. We have discussed the frequency bands of the IP and OP modes of infinite chains and the deformation of the dipole active bands due to the coupling with the surface modes of the InSb substrate and the emergence of topological edge modes for different doping levels in InSb. The coupling to the surface modes allows for a long-range heat flow through the chain in the topological trivial and non-trivial phase. The length scale of this long-range heat flow is determined by the propagation length of the surface modes. In general, if the propagation length of the surface modes is much longer than the lattice constant we find that in the non-trivial phase the edge mode contribution enhances the heat transfer through the chain compared to the trivial phase. For the case where the propagation length of the surface modes is comparable to the lattice constant this enhancement can be seen for very short chains with a length comparable to the propagation length. {Our discussion of the dependence on the chain-surface distance suggests that for certain configurations also the trivial phase can give a dominant heat flux so that for a measurement of this effect the configuration has to be chosen correspondingly to show enhancement by the edge modes. We note that the calculated power $P_N$ received by the last nanoparticle $N$ is directly associated to the conductance between the first and last particle which is $P_N/\Delta T$ where $\Delta T$ is the applied temperature difference. Furthermore, via Newtons law of cooling $P_N$ determines the temperature change $\rd T_N / \rd t$ of the last nanoparticle so that the observed effects of increased heat fluxes are, in principle, directly measurable by heat conduction or temperature measurements comparing a topological trivial and non-trivial chain. However, even if such changes are measured this only indirectly proves the impact of the edge modes because such experiments measure the spectrally integrated heat flux instead. To have clearer measurements of the presence of edge modes, we propose to carry out spectrally resolved measurements with scattering type microscopes like in Refs.~\cite{DeWilde,Jones,WengEtAl2018}. Such signals are directly related to the LDOS as visualized in Fig.~\ref{Fig:LDOS} and are, therefore, able to experimentally demonstrate the edge modes' impact. }

%
%
\section*{Acknowledgements}

The authors gratefully acknowledge financial support from the Niedersächsische Ministerium für Kultur und Wissenschaft (`DyNano'). Florian Herz acknowledges financial support by Deutsche Forschungsgemeinschaft under project number 570757245.

\end{document}